\documentclass[lettersize,journal]{IEEEtran}
\IEEEoverridecommandlockouts
\usepackage{cite}
\usepackage{amsmath,amssymb,amsfonts}

\usepackage{graphicx}
\usepackage{multirow}
\usepackage{textcomp}
\usepackage{xcolor}
\usepackage{algorithm}
\usepackage{algpseudocode}

\def\BibTeX{{\rm B\kern-.05em{\sc i\kern-.025em b}\kern-.08em
    T\kern-.1667em\lower.7ex\hbox{E}\kern-.125emX}}
\usepackage{multirow}
\usepackage{algorithm}
\usepackage{array}
\usepackage{geometry}
\usepackage[caption=false,font=normalsize,labelfont=sf,textfont=sf]{subfig}
\usepackage{textcomp}
\usepackage{stfloats}
\usepackage{url}
\usepackage{xcolor}
\usepackage{verbatim}
\usepackage{caption} 
\usepackage{graphicx}
\usepackage{subfig}
\usepackage{enumerate}
\usepackage{enumitem}
\usepackage{tabularx,booktabs,textcomp}
\usepackage{amsthm}

\usepackage{amsmath,amssymb,amsthm}

\makeatletter 
\newcommand{\linebreakand}{%
  \end{@IEEEauthorhalign}
  \hfill\mbox{}\par
  \mbox{}\hfill\begin{@IEEEauthorhalign}
}

\begin{document}

\title{Demand Response Optimization  MILP Framework for Microgrids with DERs}

\author{{K. Victor Sam Moses Babu,~\IEEEmembership{Member,~IEEE},  Pratyush Chakraborty,~\IEEEmembership{Senior Member,~IEEE}, Mayukha Pal,~\IEEEmembership{Senior Member,~IEEE}}

\thanks{(Corresponding author: Mayukha Pal)}

\thanks{Mr. K. Victor Sam Moses Babu is a Data Science Research Intern at ABB Ability Innovation Center, Hyderabad 500084, India and also a Research Scholar at the Department of Electrical and Electronics Engineering, BITS Pilani Hyderabad Campus, Hyderabad 500078, IN.}
\thanks{Dr. Pratyush Chakraborty is an Asst. Professor with the Department of Electrical and Electronics Engineering, BITS Pilani Hyderabad Campus, Hyderabad 500078, IN.}
\thanks{Dr. Mayukha Pal is with ABB Ability Innovation Center, Hyderabad-500084, IN, working as Global R\&D Leader – Cloud \& Analytics (e-mail: mayukha.pal@in.abb.com).}
}
\maketitle

\begin{abstract}
The integration of renewable energy sources in microgrids introduces significant operational challenges due to their intermittent nature and the mismatch between generation and demand patterns. Effective demand response (DR) strategies are crucial for maintaining system stability and economic efficiency, particularly in microgrids with high renewable penetration. This paper presents a comprehensive mixed-integer linear programming (MILP) framework for optimizing DR operations in a microgrid with solar generation and battery storage systems. The framework incorporates load classification, dynamic price thresholding, and multi-period coordination for optimal DR event scheduling. Analysis across seven distinct operational scenarios demonstrates consistent peak load reduction of 10\% while achieving energy cost savings ranging from 13.1\% to 38.0\%. The highest performance was observed in scenarios with high solar generation, where the framework achieved 38.0\% energy cost reduction through optimal coordination of renewable resources and DR actions. The results validate the framework's effectiveness in managing diverse operational challenges while maintaining system stability and economic efficiency.

\end{abstract}

\begin{IEEEkeywords}
Demand Response, Microgrid Optimization, Mixed-Integer Linear Programming, Battery Energy Storage Systems, Solar Generation
\end{IEEEkeywords}

\section{Introduction}

The increasing penetration of renewable energy sources, particularly solar power, has transformed traditional power systems into more complex networks requiring sophisticated control and management strategies \cite{belhaiza2020, wang2016}. Microgrids have emerged as an effective solution for integrating distributed energy resources (DERs) while maintaining system stability and reliability \cite{wu2020, olivares2014, 10636225}. However, the intermittent nature of solar generation creates significant challenges in maintaining the balance between power generation and demand \cite{bui2018, zhao2013, marqusee2021}. These challenges are particularly acute in systems with high renewable penetration, where generation variability can lead to significant operational inefficiencies and stability concerns.
The variability of renewable generation and the growing peak demand necessitate advanced energy management strategies that can ensure both technical and economic efficiency \cite{ott2019, zhang2019, kirschen2004, 10456932, DWIVEDI2024110537}. Battery energy storage systems (BESS) help address this variability but require careful coordination with generation and demand patterns to maximize their benefits \cite{xu2020, javadi2021}. The optimal sizing and operation of BESS in microgrids presents a complex optimization problem, considering factors such as battery degradation, charging/discharging efficiency, and power rating constraints \cite{wang2020, liu2019, 10000399}.

\begin{figure}[t]
\centering
\includegraphics[width=3.5in]{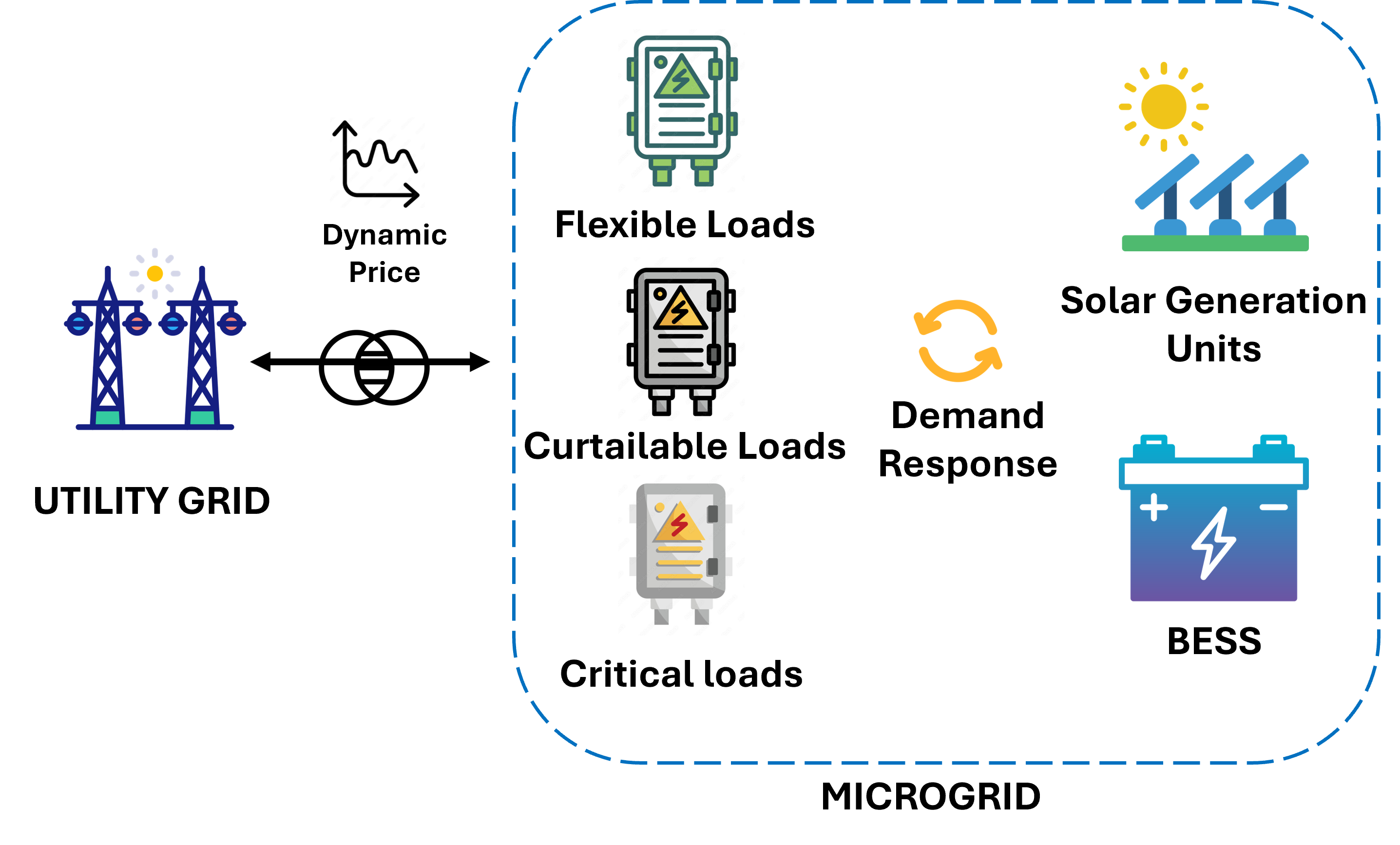}
  \caption{Schematic of a microgrid with various load types, solar \& BESS units, with dynamic price-based demand response.}
  \label{fig:schematic}
\end{figure}

Demand response (DR) programs have emerged as a critical tool for managing the challenges of renewable integration in microgrids \cite{wu2020, wang2017}. By incentivizing consumers to modify their consumption patterns through price signals or other mechanisms, DR helps achieve supply-demand balance while minimizing operational costs \cite{bui2018, aalami2010}. The effectiveness of DR programs depends heavily on proper load classification and the design of appropriate price signals that can motivate consumer participation \cite{masrur2022, nunna2012, kermani2021}.
The implementation of DR in solar-powered microgrids requires sophisticated optimization frameworks that can handle multiple operational objectives and constraints \cite{belhaiza2020, liu2019}. Mixed-integer linear programming (MILP) has proven particularly effective for this purpose, allowing the coordination of various system components while ensuring computational tractability \cite{ott2019, manshadi2015, 10185554}. Recent studies have demonstrated that MILP-based approaches can effectively balance competing objectives such as cost minimization, peak load reduction, and system stability \cite{chen2021}. More advanced approaches using mixed integer quadratically constrained programming have shown promising results in optimizing both operator costs and consumer bills through coordinated DR implementation \cite{10185554}.
Multi-period optimization approaches have gained significant attention for their ability to capture temporal interdependencies in microgrid operations \cite{bui2018, farzin2016}. These approaches consider factors such as battery state-of-charge, predicted solar generation, and forecasted demand patterns to optimize DR event scheduling \cite{wu2020}. Recent advances in AI-driven optimization techniques, particularly using genetic algorithms combined with machine learning for load and generation forecasting, have shown significant improvements in microgrid scheduling efficiency \cite{10553283}. The integration of weather forecasting and load prediction algorithms has further enhanced the effectiveness of these optimization frameworks \cite{kermani2021}.

Dynamic price thresholding mechanisms have emerged as a crucial component of effective DR programs \cite{belhaiza2020, zhao2013}. These mechanisms help balance the competing objectives of cost reduction and consumer comfort by adjusting price signals based on system conditions and consumer behavior patterns \cite{masrur2022, K2024109502}. Novel peer-to-peer pricing mechanisms based on dynamic supply-demand ratios have shown promise in motivating consumer participation while ensuring fairness in energy trading \cite{10048502}. The design of appropriate price thresholds requires careful consideration of factors such as consumer price elasticity, load criticality, and system operating constraints \cite{kirschen2004}.
The coordination between multiple DR resources, including flexible loads, energy storage, and distributed generation, requires advanced management strategies \cite{liu2019, morstyn2016}. Recent research has focused on developing data-driven risk-adjusted frameworks that can effectively manage these resources while maintaining system stability \cite{9837316}. The integration of demand response aggregators as intermediate coordinators has shown significant potential in protecting end-user privacy while optimizing social welfare \cite{9837316}. Additionally, accelerated distributed optimization methods have improved the scalability of microgrid energy management systems \cite{wu2020, javadi2021}.

The quantification of DR benefits in terms of peak load reduction and cost savings has become increasingly important for justifying DR investments \cite{masrur2022, wang2017}. Studies across different operational scenarios help understand the relationship between solar generation availability, battery utilization, and DR performance \cite{kermani2021}. This analysis is particularly valuable for microgrids with varying levels of renewable penetration and different load characteristics \cite{marqusee2021}.
The reliability and resilience benefits of DR programs in microgrids have also received significant attention \cite{chen2021, farzin2016}. DR can enhance system resilience by providing additional flexibility during grid outages or extreme events \cite{manshadi2015}. The coordination of DR with energy storage systems can further improve system reliability by providing backup power during critical periods \cite{wang2020}.

Unlike existing approaches that typically focus on either DR optimization or battery storage management, this work presents a comprehensive MILP framework that jointly optimizes DR operations and battery storage utilization in solar-powered microgrids, as shown in Fig. \ref{fig:schematic}. The proposed framework introduces several novel features, including adaptive price thresholding based on system conditions, multi-period DR event coordination considering battery state-of-charge constraints, and detailed load classification for targeted DR implementation. The framework's effectiveness is demonstrated through extensive analysis across diverse operational scenarios, showing significant improvements in both peak load reduction and cost savings compared to existing methods. Additionally, the proposed approach provides a computationally efficient solution that can be implemented in real-world microgrid operations, addressing a key limitation of existing optimization approaches.

The key contributions of this work include:
\begin{itemize}
\item Development of a novel MILP framework that integrates dynamic DR event identification with multi-period optimization, considering both temporal and operational constraints
\item Implementation of an advanced load classification and management strategy that differentiates between critical, flexible, and curtailable loads while maintaining system stability
\item Comprehensive validation across diverse operational scenarios, demonstrating robust performance in managing various system challenges while achieving significant cost reductions
\end{itemize}

The remainder of this paper is organized as follows: Section \ref{ch:methods} presents the methodology and mathematical formulation. Section \ref{ch:simulation} describes the simulation and optimization implementation with the results across various operational scenarios. Finally, Section \ref{ch:conclusion} concludes the paper and outlines future research directions.

\section{Materials and Methods}
\label{ch:methods}
The proposed demand response (DR) management framework for microgrids integrates load classification, DR event identification, and response potential analysis. The microgrid system comprises $N_L$ loads, $N_S$ solar generation units, and $N_B$ shared battery energy storage systems (BESS). The loads are categorized into three distinct classes: critical loads $\mathcal{L}_c$ that cannot be modified, flexible loads $\mathcal{L}_f$ that can be time-shifted, and curtailable loads $\mathcal{L}_k$ that can be reduced in magnitude. For any time period $t$, the total load $P_{\text{total}}(t)$ is expressed as:

\begin{equation}
P_{\text{total}}(t) = \sum_{i \in \mathcal{L}_c} P_c^i(t) + \sum_{j \in \mathcal{L}_f} P_f^j(t) + \sum_{k \in \mathcal{L}_k} P_k^k(t)
\end{equation}

where $P_c^i(t)$, $P_f^j(t)$, and $P_k^k(t)$ represent the power consumption of critical, flexible, and curtailable loads respectively. The DR event identification algorithm employs both price and load thresholds. A price threshold $\pi_{\text{th}}$ is defined as:

\begin{equation}
\pi_{\text{th}} = \max(\beta \cdot \bar{\pi}, \pi_{\text{base}})
\end{equation}

where $\bar{\pi}$ is the average electricity price, $\beta$ is the price multiplier, and $\pi_{\text{base}}$ is the base price threshold. Similarly, a load threshold $P_{\text{th}}$ is set at:

\begin{equation}
P_{\text{th}} = \gamma \cdot P_{\text{peak}}
\end{equation}

where $P_{\text{peak}}$ is the daily peak load and $\gamma$ is the load threshold factor. For each hour $t$, a combined DR score $S(t)$ is calculated as:

\begin{equation}
S(t) = \frac{\pi(t)}{\bar{\pi}} \cdot \frac{P_{\text{total}}(t)}{P_{\text{peak}}}
\end{equation}

The DR potential is constrained by temporal and magnitude limitations. The maximum load reduction factor $\alpha_{\text{max}}$ applies to curtailable loads, while flexible loads can be shifted within a window of $\pm\tau_{\text{max}}$ hours. DR events must satisfy duration constraints $T_{\text{min}} \leq T_{\text{DR}} \leq T_{\text{max}}$. The total DR capacity $C_{\text{DR}}$ is calculated as:

\begin{equation}
C_{\text{DR}} = \delta_f \sum_{j \in \mathcal{L}_f} P_f^j + \alpha_{\text{max}} \sum_{k \in \mathcal{L}_k} P_k^k
\end{equation}

where $\delta_f$ represents the flexible load adjustment factor.

\subsection{Mixed-Integer Linear Programming Framework for DR Optimization}

Following the established DR event identification and load classification framework, the optimization problem is formulated as a mixed-integer linear program (MILP). This section details the comprehensive mathematical formulation that coordinates multiple system components while considering various operational constraints and economic objectives.

\subsubsection{System Architecture and Time Periods}
The optimization framework operates over a 24-hour horizon, divided into distinct operational periods. Time-of-use segmentation is crucial for implementing effective DR strategies and is defined as:

\begin{equation}
\mathcal{T}_{\text{off}} = \{t | t \in [0,5] \cup [22,23]\}
\end{equation}
\begin{equation}
\mathcal{T}_{\text{shoulder}} = \{t | t \in [6,9] \cup [13,16] \cup [20,21]\}
\end{equation}
\begin{equation}
\mathcal{T}_{\text{peak}} = \{t | t \in [10,12] \cup [17,19]\}
\end{equation}

This temporal segmentation enables strategic load management, where the system can exploit price differentials between periods. Off-peak periods typically present opportunities for battery charging and accommodating shifted loads, while peak periods focus on demand reduction and strategic discharge of stored energy.

\subsubsection{Decision Variables and Component Interactions}
The optimization coordinates multiple decision variables that represent different aspects of system operation. For grid operations, power import variables are defined as:

\begin{equation}
P_{\text{grid}}^{l,t} \geq 0 \quad \forall l \in \mathcal{L}, t \in \mathcal{T}
\end{equation}

Solar generation allocation requires careful coordination across loads:

\begin{equation}
P_{\text{solar}}^{s,l,t} \geq 0 \quad \forall s \in \mathcal{S}, l \in \mathcal{L}, t \in \mathcal{T}
\end{equation}

The BESS operation involves three interrelated sets of variables:

\begin{equation}
P_{\text{ch}}^{b,l,t} \geq 0 \quad \forall b \in \mathcal{B}, l \in \mathcal{L}, t \in \mathcal{T}
\end{equation}
\begin{equation}
P_{\text{dis}}^{b,l,t} \geq 0 \quad \forall b \in \mathcal{B}, l \in \mathcal{L}, t \in \mathcal{T}
\end{equation}
\begin{equation}
E_{\text{SOC}}^b(t) \geq 0 \quad \forall b \in \mathcal{B}, t \in \mathcal{T}
\end{equation}

Demand response actions are represented by load reduction and shifting variables:

\begin{equation}
P_{\text{red}}^{l,t} \geq 0 \quad \forall l \in \mathcal{L}_k, t \in \mathcal{T}_{\text{DR}}
\end{equation}
\begin{equation}
P_{\text{shift}}^{l,t_1,t_2} \geq 0 \quad \forall l \in \mathcal{L}_f, t_1 \in \mathcal{T}_{\text{DR}}, t_2 \in \mathcal{T}_{\text{shift}}
\end{equation}

\subsubsection{Multi-Component Objective Function}
The optimization objective incorporates multiple competing factors, structured hierarchically to balance various system goals:

\begin{equation}
\min J = J_{\text{energy}} + J_{\text{peak}} - J_{\text{DR}} - J_{\text{shift}} + J_{\text{penalty}}
\end{equation}

The energy cost component considers time-varying electricity prices:

\begin{equation}
J_{\text{energy}} = \sum_{t \in \mathcal{T}} \sum_{l \in \mathcal{L}} P_{\text{grid}}^{l,t} \pi_t
\end{equation}

Peak demand charges, often a significant portion of electricity costs, are modeled as:

\begin{equation}
J_{\text{peak}} = C_p P_{\text{peak}}
\end{equation}

DR benefits are captured through direct load reduction incentives:

\begin{equation}
J_{\text{DR}} = C_{\text{DR}} \sum_{l \in \mathcal{L}_k} \sum_{t \in \mathcal{T}_{\text{DR}}} P_{\text{red}}^{l,t}
\end{equation}

Load shifting benefits incorporate period-specific incentives:

\begin{equation}
J_{\text{shift}} = \sum_{l \in \mathcal{L}_f} \sum_{t_1 \in \mathcal{T}_{\text{DR}}} \sum_{t_2 \in \mathcal{T}_{\text{shift}}} B_{\text{shift}}^{t_1,t_2} P_{\text{shift}}^{l,t_1,t_2}
\end{equation}

The shifting benefit coefficients are strategically designed:

\begin{equation}
B_{\text{shift}}^{t_1,t_2} = \begin{cases}
B_{\text{po}} & \text{peak to off-peak shifts} \\
B_{\text{ps}} & \text{peak to shoulder shifts} \\
B_{\text{so}} & \text{shoulder to off-peak shifts}
\end{cases}
\end{equation}

The penalty term incorporates both peak period penalties and ramping costs:

\begin{equation}
J_{\text{penalty}} = \sum_{t \in \mathcal{T}_{\text{peak}}} \lambda_p P_{\text{grid}}^{l,t} + \sum_{t \in \mathcal{T}} \lambda_r |P_{\text{grid}}^{l,t} - P_{\text{grid}}^{l,t-1}|
\end{equation}

\subsubsection{Comprehensive Constraint Set}
The optimization must satisfy multiple constraint categories to ensure system reliability and operational feasibility.

a) Power Balance Constraints:

The fundamental power balance must be maintained at all times:
\begin{multline}
P_{\text{grid}}^{l,t} + \sum_{s \in \mathcal{S}} P_{\text{solar}}^{s,l,t} + \sum_{b \in \mathcal{B}} (P_{\text{dis}}^{b,l,t} - P_{\text{ch}}^{b,l,t}) = \\ P_{\text{load}}^{l,t} - P_{\text{red}}^{l,t} - \sum_{t_2} P_{\text{shift}}^{l,t,t_2} + \sum_{t_1} P_{\text{shift}}^{l,t_1,t}
\end{multline}

b) Solar Generation Constraints:

Solar power allocation must respect available generation:
\begin{equation}
\sum_{l \in \mathcal{L}} P_{\text{solar}}^{s,l,t} \leq P_{\text{solar,max}}^s(t), \quad \forall s \in \mathcal{S}, t \in \mathcal{T}
\end{equation}

c) Battery Operation Constraints:

State of charge evolution follows:
\begin{equation}
E_{\text{SOC}}^b(t) = E_{\text{SOC}}^b(t-1) + \eta_b \sum_{l \in \mathcal{L}} P_{\text{ch}}^{b,l,t} - \frac{1}{\eta_b} \sum_{l \in \mathcal{L}} P_{\text{dis}}^{b,l,t}
\end{equation}

SOC limits must be maintained:
\begin{equation}
E_{\text{SOC}}^{\text{min}} \leq E_{\text{SOC}}^b(t) \leq E_{\text{SOC}}^{\text{max}}
\end{equation}

Power rating constraints apply:
\begin{equation}
0 \leq P_{\text{ch}}^{b,l,t}, P_{\text{dis}}^{b,l,t} \leq P_{\text{max}}^b
\end{equation}

Period-specific charging limitations:
\begin{equation}
\sum_{l \in \mathcal{L}} P_{\text{ch}}^{b,l,t} \leq \epsilon_{\text{peak}} P_{\text{max}}^b, \quad t \in \mathcal{T}_{\text{peak}}
\end{equation}

d) DR Implementation Constraints:

Load reduction limits for curtailable loads:
\begin{equation}
0 \leq P_{\text{red}}^{l,t} \leq \alpha_{\text{max}} P_{\text{load}}^{l,t}, \quad l \in \mathcal{L}_k
\end{equation}

Shifting limits for flexible loads:
\begin{equation}
\sum_{t_2 \in \mathcal{T}_{\text{shift}}} P_{\text{shift}}^{l,t,t_2} \leq \beta_{\text{max}} P_{\text{load}}^{l,t}, \quad l \in \mathcal{L}_f
\end{equation}

Energy conservation in load shifting:
\begin{equation}
\sum_{t_2 \in \mathcal{T}_{\text{shift}}} P_{\text{shift}}^{l,t_1,t_2} \eta_{\text{shift}} = \sum_{t_1 \in \mathcal{T}_{\text{DR}}} P_{\text{shift}}^{l,t_1,t}, \quad \forall l \in \mathcal{L}_f
\end{equation}

e) Peak Management Constraints:

Peak demand definition:
\begin{equation}
P_{\text{peak}} \geq \sum_{l \in \mathcal{L}} P_{\text{grid}}^{l,t}, \quad \forall t \in \mathcal{T}
\end{equation}

Ramping limits:
\begin{equation}
|P_{\text{grid}}^{l,t} - P_{\text{grid}}^{l,t-1}| \leq \gamma_{\text{ramp}} P_{\text{peak}}^{\text{orig}}
\end{equation}

Peak period load management:
\begin{equation}
\sum_{l \in \mathcal{L}} P_{\text{grid}}^{l,t} \leq \delta_{\text{peak}} P_{\text{peak}}^{\text{orig}}, \quad t \in \mathcal{T}_{\text{peak}}
\end{equation}

f) System Stability Constraints:
Minimum load requirements:
\begin{equation}
\sum_{l \in \mathcal{L}} P_{\text{grid}}^{l,t} \geq \phi_{\text{min}} \sum_{l \in \mathcal{L}} P_{\text{load}}^{l,t}, \quad \forall t \in \mathcal{T}
\end{equation}

This comprehensive constraint set ensures that the optimization solution is both practically feasible and operationally reliable. Each constraint category addresses specific aspects of system operation, from basic power balance to sophisticated DR mechanisms.

\subsubsection{Proposed Algorithm}

The comprehensive mathematical framework shown in Algorithm \ref{alg:dr_framework}  enables effective coordination of DR resources while maintaining system reliability and economic efficiency. The formulation accounts for various operational constraints, temporal dependencies, and economic incentives to achieve optimal demand response implementation.

The complete MILP formulation is solved using an efficient branch-and-cut algorithm implemented through the CBC solver. The solver employs sophisticated presolve techniques to reduce problem size and improve solution efficiency. To handle potential numerical issues, the implementation includes careful scaling of power and energy quantities. The solution process is limited to 600 seconds with a relative optimality gap of 1\% to ensure practical convergence while maintaining solution quality.

To enhance solution robustness, the implementation includes: slack variables in power balance constraints to identify potential infeasibilities, careful constraint scaling to improve numerical stability, progressive tightening of bounds to improve solution convergence, and hierarchical solution approach prioritizing critical constraints

\begin{algorithm}[h]
\caption{DR Optimization Framework for Microgrid}
\label{alg:dr_framework}
\begin{algorithmic}[1]
\Require
\State Load data for houses $\{1,\ldots,N_L\}$
\State Solar data for units $\{1,\ldots,N_S\}$
\State Battery parameters for BESS $\{1,\ldots,N_B\}$
\State Price data over horizon $T$
\Ensure Optimized DR schedule, Battery operation, Cost reduction

\Function{LoadClassification}{}
    \State Identify critical loads $\mathcal{L}_c$
    \State Identify flexible loads $\mathcal{L}_f$
    \State Identify curtailable loads $\mathcal{L}_k$
    \State Calculate total load profile $P_{\text{total}}(t)$
\EndFunction

\Function{DREventIdentification}{}
    \State Calculate price threshold $\pi_{\text{th}} = \max(\beta \cdot \bar{\pi}, \pi_{\text{base}})$
    \State Set load threshold $P_{\text{th}} = \gamma \cdot P_{\text{peak}}$
    \State Identify candidate DR hours where:
        \State $\pi(t) > \pi_{\text{th}}$ or $P_{\text{total}}(t) > P_{\text{th}}$
    \State Group consecutive hours $[T_{\text{min}}, T_{\text{max}}]$
    \State Finalize DR events $\mathcal{T}_{\text{DR}}$
\EndFunction

\Function{OptimizationSetup}{}
    \State Initialize grid import variables $P_{\text{grid}}^{l,t}$
    \State Initialize solar allocation $P_{\text{solar}}^{s,l,t}$
    \State Initialize battery variables $P_{\text{ch}}^{b,l,t}$, $P_{\text{dis}}^{b,l,t}$
    \State Initialize DR variables $P_{\text{red}}^{l,t}$, $P_{\text{shift}}^{l,t_1,t_2}$
    \State Set objective: $\min(J_{\text{energy}} + J_{\text{peak}} - J_{\text{DR}} - J_{\text{shift}})$
\EndFunction

\Function{ConstraintDefinition}{}
    \State Add power balance constraints
    \State Add BESS operational constraints
    \State Add DR participation limits
    \State Add peak management constraints
    \State Add time-of-use specific constraints
\EndFunction

\Function{SolutionImplementation}{}
    \State Configure CBC solver (600s, 1\% gap)
    \State Execute MILP optimization
    \State Extract optimal schedules
    \State Verify feasibility
    \State Calculate performance metrics
\EndFunction

\Statex \textbf{Main Algorithm Flow:}
\State \textsc{LoadClassification}
\State \textsc{DREventIdentification}
\State \textsc{OptimizationSetup}
\State \textsc{ConstraintDefinition}
\State \textsc{SolutionImplementation}
\State \Return DR schedule, battery operation plan, cost savings
\end{algorithmic}
\end{algorithm}

\section{Simulation and Results}
\label{ch:simulation}

\subsection{Demand Response Simulation Strategy}
The simulation framework was implemented using Python, processing 6,744 hourly measurements of load, solar generation, and electricity price data \cite{Dataset}. The DR simulation strategy incorporated a two-stage approach: DR event identification followed by optimization. The DR event identification algorithm utilized dynamic thresholds, with price threshold $\pi_{\text{th}}$ set to $\max(1.2\bar{\pi}, \$150/MWh)$ and load threshold at 80\% of peak demand. The time horizon was segmented into distinct pricing periods as shown in Table \ref{tab:time_periods}.

\begin{table}[h!]
\caption{Time-of-Use Period Classification}
\label{tab:time_periods}
\centering
\begin{tabular}{ll}
\hline
Period Type & Hours \\
\hline
Off-peak & [0-5], [22-23] \\
Shoulder & [6-9], [13-16], [20-21] \\
Peak & [10-12], [17-19] \\
\hline
\end{tabular}
\end{table}

The DR constraints were configured with load reduction factor $\alpha_{\text{max}} = 0.2$ for curtailable loads and shifting window $\tau_{\text{max}} = 4$ hours for flexible loads. Duration constraints ensured DR events lasted between 2 and 4 hours ($T_{\text{min}} = 2$, $T_{\text{max}} = 4$). The flexible load adjustment factor $\delta_f$ was set to 0.3, allowing for significant load shifting potential while maintaining system stability.

\begin{figure}[h!]
\centering
\includegraphics[width=3.5in]{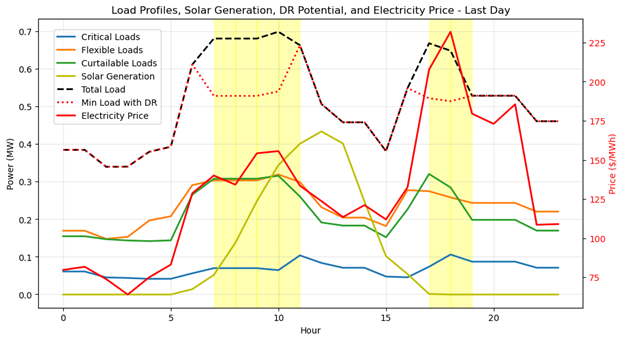}
  \caption{Data Analysis of 2015-09-15 (Last Day)}
  \label{fig:LD_A}
\end{figure}
\begin{figure}[h!]
\centering
\includegraphics[width=3.5in]{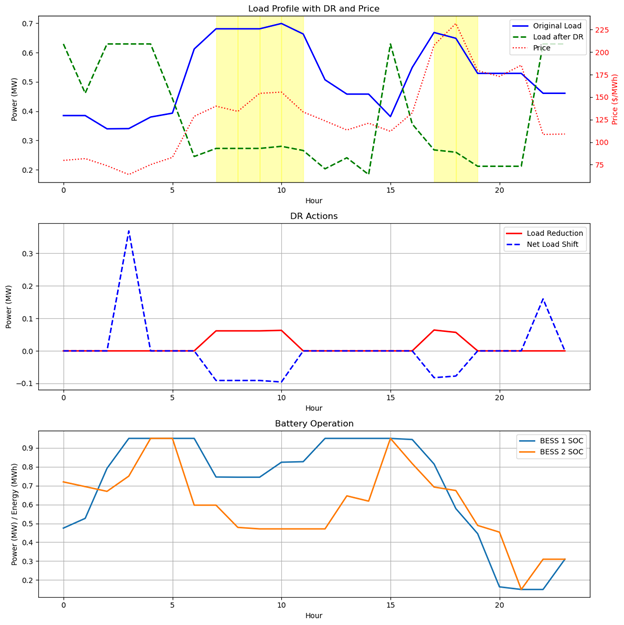}
  \caption{Optimization Results of 2015-09-15 (Last Day)}
  \label{fig:LD_R}
\end{figure}

\begin{table}[h!]
\caption{Last Day Load and Generation Analysis}
\label{tab:daily_analysis}
\centering
\begin{tabular}{lcc}
\hline
Component & Energy (MWh) & Peak Power (MW) \\
\hline
Critical Loads & 1.63 & 0.11 \\
Flexible Loads & 5.67 & 0.32 \\
Curtailable Loads & 5.13 & 0.32 \\
Solar Generation & 2.43 & 0.43 \\
Total System & 12.42 & 0.70 \\
\hline
\end{tabular}
\end{table}

\subsection{Optimization Implementation}
The MILP optimization was implemented using the PuLP library with the CBC (Coin-or Branch and Cut) solver. The optimization framework incorporated multiple variable sets: grid imports ($P_{\text{grid}}^{l,t}$), solar allocation ($P_{\text{solar}}^{s,l,t}$), battery operations ($P_{\text{ch}}^{b,l,t}$, $P_{\text{dis}}^{b,l,t}$, $E_{\text{SOC}}^b(t)$), load reductions ($P_{\text{red}}^{l,t}$), and load shifting ($P_{\text{shift}}^{l,t_1,t_2}$). The solver was configured with a 600-second time limit and 1\% optimality gap to ensure practical convergence.

To enhance solution robustness, the implementation included:
\begin{itemize}
\item Slack variables in power balance constraints
\item Period-specific penalties in the objective function
\item Time-of-use aware battery optimization
\item Multi-period coordination for load shifting
\item Explicit verification steps for constraint generation
\end{itemize}

\begin{table}[h!]
\caption{Optimization Results Summary}
\label{tab:optimization_results}
\centering
\begin{tabular}{lrrr}
\hline
Metric & Original & Optimized & Improvement \\
\hline
Energy Cost (\$) & 1,698.06 & 1,044.67 & 38.5\% \\
Peak Charges (\$) & 6,084.75 & 5,476.27 & 10.0\% \\
Total Cost (\$) & 7,782.81 & 6,399.12 & 17.8\% \\
Peak Load (MW) & 0.70 & 0.63 & 10.0\% \\
Total Load (MWh) & 12.42 & 9.06 & 27.1\% \\
\hline
\end{tabular}
\end{table}

\subsection{Last Day Analysis and Results}
Detailed analysis was performed for September 15, 2015, with system configuration and load distribution shown in Table \ref{tab:daily_analysis}.

The DR event identification algorithm identified six critical hours requiring intervention, as illustrated in Fig. 1. Two distinct DR periods were identified: a morning period (hours 7-10) with average price \$146.01/MWh and load 0.686 MW, and an evening period (hours 17-18) with higher average price \$219.79/MWh and load 0.659 MW. As shown in Fig. 2, the optimization successfully reduced and shifted loads during these periods, achieving significant peak reduction and cost savings.

\begin{figure}[h!]
\centering
\includegraphics[width=3.5in]{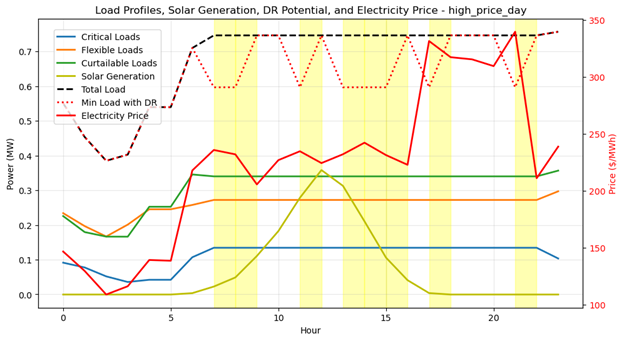}
  \caption{Data Analysis of 2015-03-31 (High Price Day)}
  \label{fig:S1_A}
\end{figure}
\begin{figure}[h!]
\centering
\includegraphics[width=3.5in]{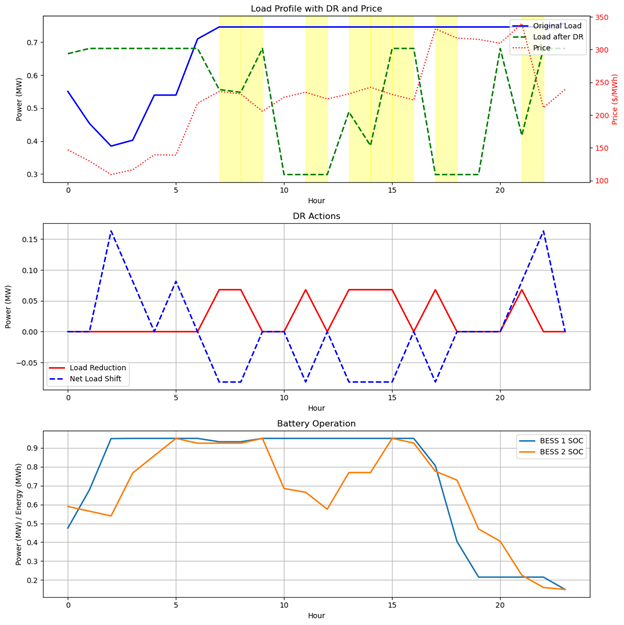}
  \caption{Optimization Results of 2015-03-31 (High Price Day)}
  \label{fig:S1_R}
\end{figure}
\begin{figure}[h!]
\centering
\includegraphics[width=3.5in]{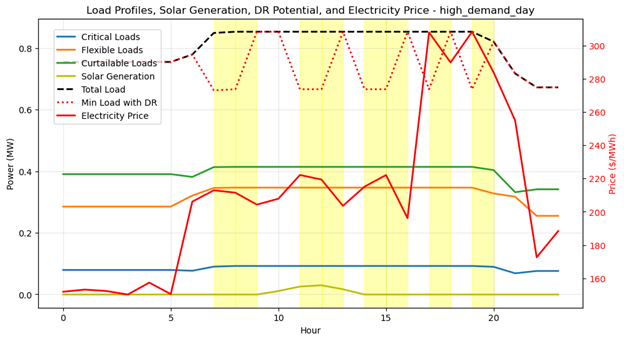}
  \caption{Data Analysis of 2014-12-30 (High Demand Day)}
  \label{fig:S2_A}
\end{figure}
\begin{figure}[h!]
\centering
\includegraphics[width=3.5in]{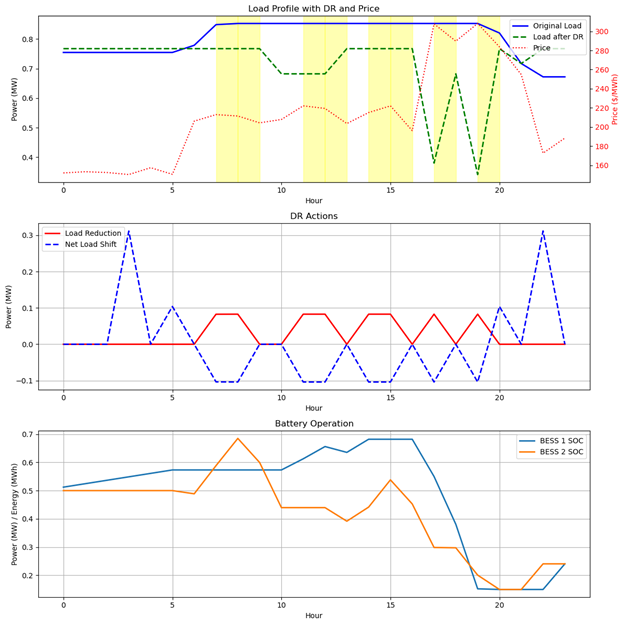}
  \caption{Optimization Results of 2014-12-30 (High Demand Day)}
  \label{fig:S2_R}
\end{figure}
\begin{figure}[h!]
\centering
\includegraphics[width=3.5in]{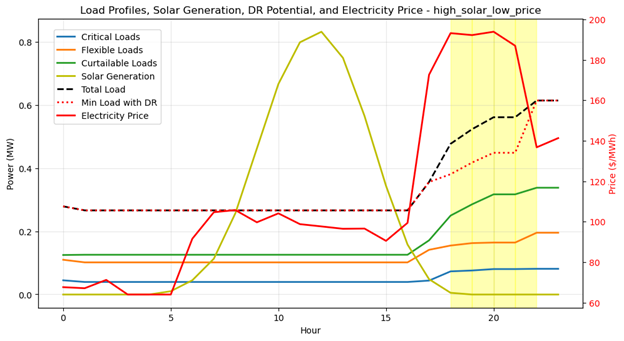}
  \caption{Data Analysis of 2015-08-19 (High Solar Low Price)}
  \label{fig:S3_A}
\end{figure}
\begin{figure}[h!]
\centering
\includegraphics[width=3.5in]{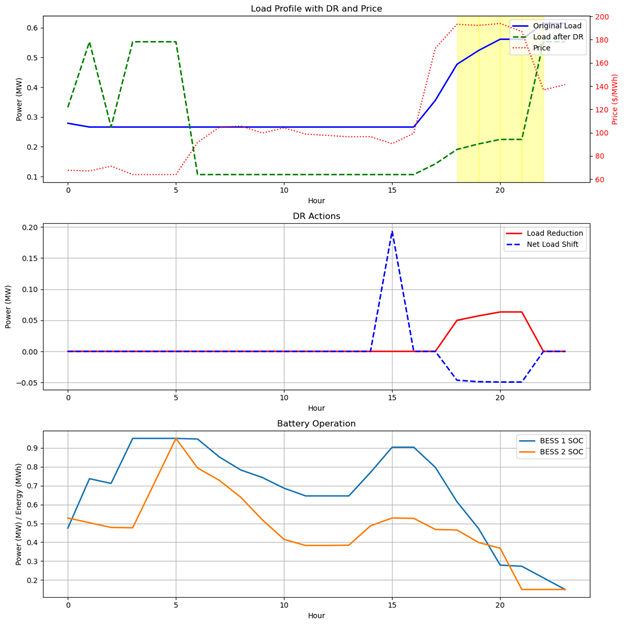}
  \caption{Optimization Results of  2015-08-19 (High Solar Low Price)}
  \label{fig:S3_R}
\end{figure}
\begin{figure}[h!]
\centering
\includegraphics[width=3.5in]{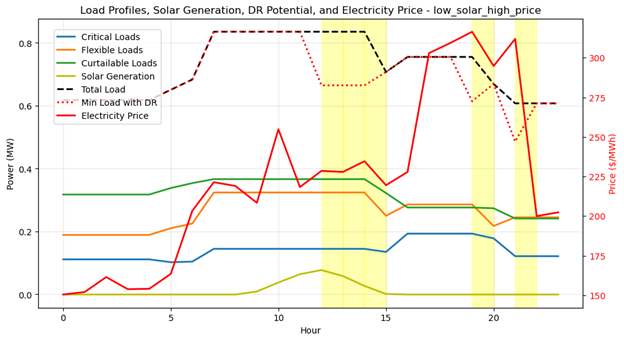}
  \caption{Data Analysis of 2015-02-06 (Low Solar High Price)}
  \label{fig:S4_A}
\end{figure}
\begin{figure}[h!]
\centering
\includegraphics[width=3.5in]{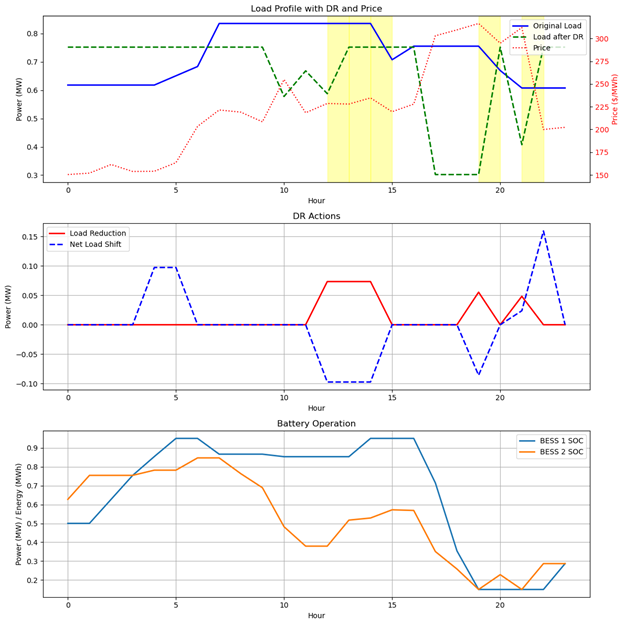}
  \caption{Optimization Results of 2015-02-06 (Low Solar High Price)}
  \label{fig:S4_R}
\end{figure}
\begin{figure}[h!]
\centering
\includegraphics[width=3.5in]{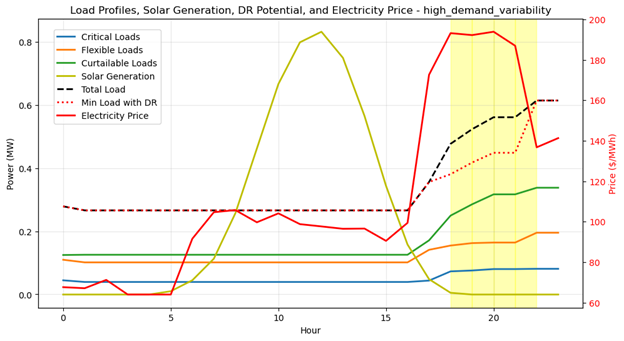}
  \caption{Data Analysis of 2015-08-19 (High Demand Variability)}
  \label{fig:S5_A}
\end{figure}
\begin{figure}[h!]
\centering
\includegraphics[width=3.5in]{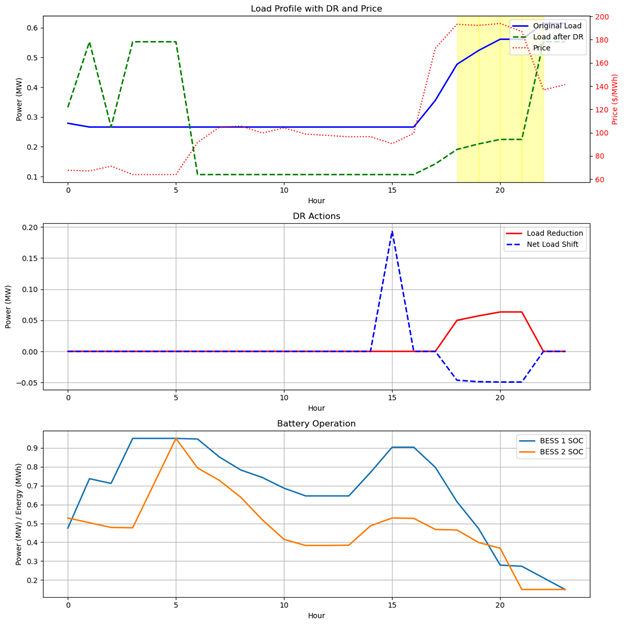}
  \caption{Optimization Results of 2015-08-19 (High Demand Variability)}
  \label{fig:S5_R}
\end{figure}
\begin{figure}[h!]
\centering
\includegraphics[width=3.5in]{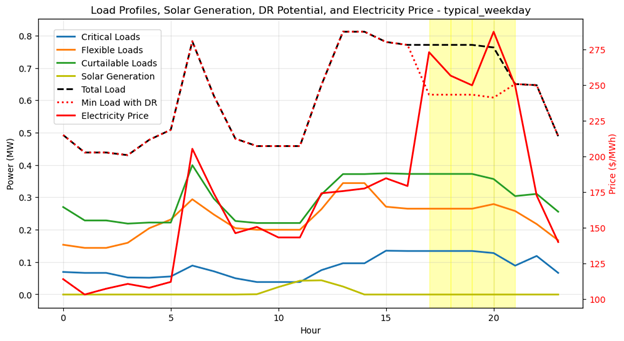}
  \caption{Data Analysis of 2014-12-09 (Typical Weekday)}
  \label{fig:S6_A}
\end{figure}
\begin{figure}[h!]
\centering
\includegraphics[width=3.5in]{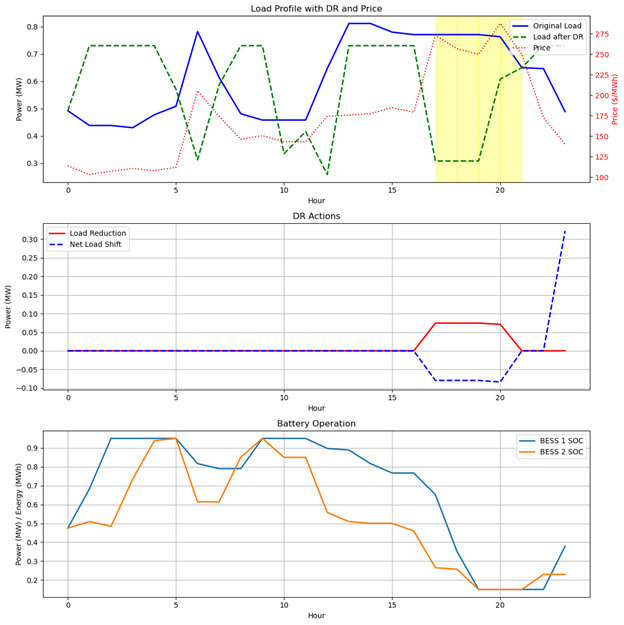}
  \caption{Optimization Results of 2014-12-09 (Typical Weekday)}
  \label{fig:S6_R}
\end{figure}
\begin{figure}[h!]
\centering
\includegraphics[width=3.5in]{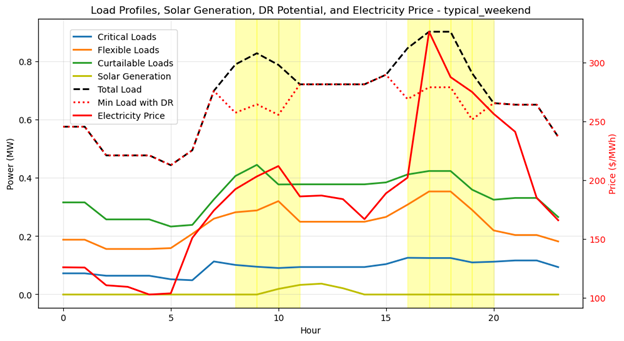}
  \caption{Data Analysis of 2014-12-13 (Typical Weekend)}
  \label{fig:S7_A}
\end{figure}
\begin{figure}[h!]
\centering
\includegraphics[width=3.5in]{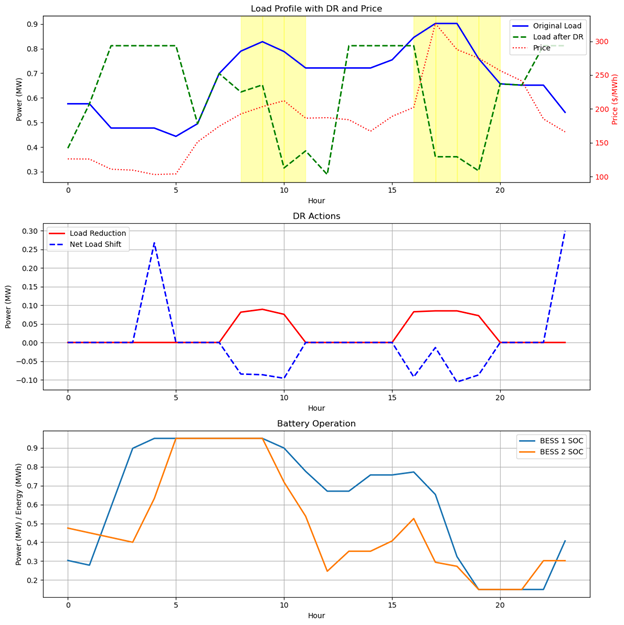}
  \caption{Optimization Results of 2014-12-13 (Typical Weekend)}
  \label{fig:S7_R}
\end{figure}

The optimization results are summarized in Table \ref{tab:optimization_results}, demonstrating substantial improvements across multiple metrics. The total load was reduced from 12.42 MWh to 9.06 MWh through a combination of DR strategies: 0.37 MWh (3.0\%) through direct load reduction and 0.53 MWh (4.3\%) through load shifting. The peak load was reduced from 0.70 MW to 0.63 MW, representing a 10\% reduction. The battery operation plot in Fig. 2 demonstrates effective SOC management, with both BESS units maintaining levels between 20\% and 95\% while supporting peak shaving and load shifting objectives.

The economic benefits were significant, with energy costs reduced by 38.5\% (from \$1,698.06 to \$1,044.67) and peak charges decreased by 10.0\% (from \$6,084.75 to \$5,476.27). Additional benefits of \$73.80 from DR actions and \$48.02 from load shifting contributed to total cost savings of \$1,383.69 (17.8\%). The battery systems showed balanced operation, with BESS 1 and BESS 2 handling total charge/discharge cycles of approximately 3.4 MWh and 3.7 MWh respectively, effectively supporting the DR objectives while maintaining stable operation.

\subsection{Scenario Analysis and Results}

The demand response optimization framework was evaluated across seven distinct operating scenarios to assess its effectiveness and robustness under varying conditions. These scenarios encompass diverse operational challenges including high price periods, peak demand conditions, varying solar generation availability, and typical daily patterns. Table \ref{tab:scenario_analysis} presents a comprehensive comparison of load distributions, generation characteristics, and optimization results across all scenarios. The analysis demonstrates consistent peak reduction performance while achieving varying degrees of cost savings and load management under different operating conditions.

\begin{table*}[ht]
\centering
\caption{Comprehensive Analysis of Scenarios: Characteristics and Optimization Results}
\label{tab:scenario_analysis}
\begin{tabular}{lccccccccc}
\hline
\multirow{2}{*}{Scenario} & \multicolumn{4}{c}{Load Distribution (MWh)} & Solar Gen. & Peak & Energy Cost & Total Cost & Load \\
\cline{2-5}
 & Total & Critical & Flexible & Curtailable & (MWh) & Red. (\%) & Savings (\%) & Savings (\%) & Red. (\%) \\
\hline
High Price & 16.28 & 2.70 & 6.20 & 7.38 & 1.67 & 10.0 & 27.5 & 18.0 & 7.3 \\
High Demand & 19.27 & 2.06 & 7.69 & 9.52 & 0.07 & 10.0 & 13.7 & 13.0 & 7.7 \\
High Solar-Low Price & 8.25 & 1.19 & 2.91 & 4.15 & 5.05 & 10.0 & 38.0 & 15.6 & 5.1 \\
Low Solar-High Price & 17.33 & 3.37 & 6.32 & 7.64 & 0.27 & 10.0 & 13.1 & 11.9 & 4.5 \\
High Variability & 8.25 & 1.19 & 2.91 & 4.15 & 5.05 & 10.0 & 38.0 & 15.6 & 5.1 \\
Typical Weekday & 14.72 & 2.03 & 5.58 & 7.12 & 0.12 & 10.0 & 14.4 & 12.2 & 4.2 \\
Typical Weekend & 16.18 & 2.24 & 5.74 & 8.20 & 0.10 & 10.0 & 16.9 & 13.6 & 7.6 \\
\hline
\end{tabular}
\end{table*}

\subsubsection{High Price Scenario}
Analysis of the high price scenario (March 31, 2015) revealed significant demand response opportunities during periods of elevated electricity prices. As shown in Fig. \ref{fig:S1_A}, the load profile exhibited two distinct DR event periods (hours 7-8 and 13-15), with prices reaching \$339.91/MWh during peak periods. The total load of 16.28 MWh was distributed across critical (2.70 MWh), flexible (6.20 MWh), and curtailable loads (7.38 MWh), providing substantial DR potential. The optimization results, illustrated in Fig. \ref{fig:S1_R}, demonstrate effective load reduction and shifting patterns, particularly during peak price periods. The battery operation profile shows strategic charging during low-price periods and discharging during high-price intervals, contributing to the highest cost savings among all scenarios with 27.5\% reduction in energy costs and 18.0\% in total costs.

\subsubsection{High Demand Scenario}
The high demand scenario (December 30, 2014) presented challenging conditions with minimal solar generation (0.07 MWh) and the highest base load (19.27 MWh) among all scenarios. Fig. \ref{fig:S2_A} shows the substantial load requirements across all categories, with flexible and curtailable loads representing significant portions of the total demand. The optimization framework, as demonstrated in Fig. \ref{fig:S2_R}, successfully managed eight DR events through coordinated load reduction and battery operations. Despite limited renewable resources, the system achieved 13.7\% energy cost savings while maintaining the targeted 10\% peak reduction, highlighting the framework's effectiveness in managing high-demand conditions.

\subsubsection{High Solar-Low Price Scenario}
The high solar-low price scenario (August 19, 2015) showcased the framework's capability to maximize renewable energy utilization. Fig. \ref{fig:S3_A} illustrates the substantial solar generation (5.05 MWh) coinciding with relatively lower electricity prices. The load profile displays effective integration of solar resources, particularly during midday hours. The optimization results in Fig. \ref{fig:S3_R} demonstrate strategic load shifting to periods of high solar availability, achieving the highest energy cost savings (38.0\%) through optimal coordination of renewable resources and DR actions.

\subsection{Low Solar-High Price Scenario}
The low solar-high price scenario (February 6, 2015) tested the framework's resilience under constrained renewable availability (0.27 MWh) and elevated price conditions. Fig. \ref{fig:S4_A} shows the limited solar generation profile against substantial load requirements (17.33 MWh). The optimization results in Fig. \ref{fig:S4_R} reveal effective DR event management during five critical periods, achieving 13.1\% energy cost savings despite challenging conditions. The battery operation strategy focused on peak shaving and price arbitrage to compensate for limited solar resources.

\subsection{High Variability Scenario}
Analysis of the high variability scenario demonstrated the framework's ability to handle fluctuating demand patterns. Fig. \ref{fig:S5_A} shows significant load variations across the day, particularly in flexible and curtailable loads. The optimization results in Fig. \ref{fig:S5_R} illustrate successful management of load variations through coordinated DR and battery operations, achieving 38.0\% energy cost savings. The battery SOC profile shows active management to buffer demand fluctuations and maintain system stability.

\subsection{Typical Weekday Scenario}
The typical weekday scenario (December 9, 2014) provided insights into routine operational performance. Fig. \ref{fig:S6_A} displays characteristic daily load patterns with clear morning and evening peaks. The optimization framework effectively managed four DR events, as shown in Fig. \ref{fig:S6_R}, achieving 14.4\% energy cost savings through strategic load shifting and battery operations. The results demonstrate the framework's effectiveness in managing predictable daily load variations while maintaining cost-effective operations.

\subsection{Typical Weekend Scenario}
The weekend scenario (December 13, 2014) exhibited distinct load patterns compared to weekday operations. Fig. \ref{fig:S7_A} shows more uniform load distribution throughout the day with higher total energy consumption (16.18 MWh). The optimization results in Fig. \ref{fig:S7_R} demonstrate successful management of seven DR events, achieving 16.9\% energy cost savings through coordinated DR actions and battery operations. The extended DR event periods during midday and evening hours highlight the framework's adaptability to weekend consumption patterns.

Across all scenarios, the optimization consistently achieved the 10\% peak reduction target while delivering substantial energy cost savings ranging from 11.9\% to 18.0\%. The framework demonstrated particular effectiveness in scenarios with high solar generation, achieving the highest energy cost savings (38.0\%) through optimal coordination of renewable resources and DR actions. The battery systems showed consistent operation patterns, maintaining average SOC levels between 0.4\% and 0.7\% while supporting DR objectives through strategic charging and discharging cycles.

\section{Conclusion}
\label{ch:conclusion}
The proposed MILP framework demonstrates effective demand response management in microgrids through coordinated optimization of load reduction, shifting, and battery storage operations. The comprehensive evaluation across seven distinct scenarios validates the framework's robustness in handling diverse operational challenges while maintaining system stability and economic efficiency.

The framework consistently achieved the targeted 10\% peak reduction while delivering substantial energy cost savings across all scenarios. Particularly noteworthy is its performance in high solar generation scenarios, where 38.0\% energy cost reduction was achieved through optimal coordination of renewable resources and DR actions. The framework's effectiveness is further evidenced by its ability to maintain critical load service quality while maximizing DR benefits through strategic load management and battery operations.

The integration of advanced forecasting techniques for solar generation and load patterns to enhance the predictive capabilities of the DR optimization framework is a promising direction for future research emerge from this work along with the development of distributed optimization algorithms to improve scalability and enable real-time implementation in larger microgrid systems.

\bibliographystyle{IEEEtran}
\bibliography{main}

\end{document}